\documentclass[11pt,twoside]{article}  % Leave intact
\usepackage{adassconf}

\begin{document}   % Leave intact

%-----------------------------------------------------------------------
%			    Paper ID Code
%-----------------------------------------------------------------------
% Enter the proper paper identification code.  The ID code for your
% paper is the session number associated with your presentation as
% published in the official conference proceedings.  You can
% find this number locating your abstract in the printed proceedings
% that you received at the meeting or on-line at the conference web
% site; the ID code is the letter/number sequence proceeding the title 
% of your presentation.  
%
% This will not appear in your paper; however, it allows different
% papers in the proceedings to cross-reference each other.  Note that
% you should only have one \paperID, and it should not include a
% trailing period.
%

\paperID{O5-2}

%-----------------------------------------------------------------------
%		            Paper Title 
%-----------------------------------------------------------------------
% Enter the title of the paper.
%
% EXAMPLE: \title{A Breakthrough in Astronomical Software Development}

\title{Application of Medical Imaging Software to 3D Visualization of Astronomical Data}
%\titlemark{ }

%-----------------------------------------------------------------------
%		          Authors of Paper
%-----------------------------------------------------------------------
% Enter the authors followed by their affiliations.  The \author and
% \affil commands may appear multiple times as necessary.  List each
% author by giving the first name or initials first followed by the
% last name.  Authors with the same affiliations should grouped
% together. 
%
% Try to limit the front matter to no more than three \author
% commands.  Group authors with the same affiliations.  Too many
% \author commands fills the first page of the paper with little
% actual text.

\author{Michelle Borkin, Alyssa Goodman\altaffilmark{1}, Michael Halle\altaffilmark{2}, \& Douglas Alan}
\affil{Initiative in Innovative Computing, Harvard University, 60 Oxford Street, Cambridge MA 02138}

\altaffiltext{1}{Also affiliated with the Harvard-Smithsonian Center for Astrophysics.}
\altaffiltext{1}{Also affiliated with the Surgical Planning Lab at Brigham and Women's Hospital \& Harvard Medical School.}

%-----------------------------------------------------------------------
%			 Contact Information
%-----------------------------------------------------------------------
% This information will not appear in the paper but will be used by
% the editors in case you need to be contacted concerning your
% submission.  Enter your name as the contact along with your email
% address.

\contact{Michelle Borkin}
\email{michelle_borkin@harvard.edu }

%-----------------------------------------------------------------------
%		      Author Index Specification
%-----------------------------------------------------------------------
% Specify how each author name should appear in the author index.  The 
% \paindex{ } should be used to indicate the primary author, and the
% \aindex for all other co-authors.  You MUST use the following
% syntax: 
%
% SYNTAX:  \aindex{LASTNAME, F. M.}
% 
% where F is the first initial and M is the second initial (if
% used).  This guarantees that authors that appear in multiple papers
% will appear only once in the author index.  
%
% EXAMPLE: \paindex{Crabtree, D.}
%          \aindex{Manset, N.}
%          \aindex{Veillet, C.}
%
% NOTE: this information is also used to build the author list that
% appears in the table of contents.  Authors will be listed in the order
% of the \paindex and \aindex commmands.
%

\paindex{Borkin, M.}
\aindex{Goodman, A.}     
\aindex{Halle, M.}     
\aindex{Alan, D.}     

%-----------------------------------------------------------------------
%                     Author list for page header
%-----------------------------------------------------------------------
% Please supply a list of author last names for the page header. in
% one of these formats:
%
% EXAMPLES:
% \authormark{LASTNAME}
% \authormark{LASTNAME1 \& LASTNAME2}
% \authormark{LASTNAME1, LASTNAME2, ... \& LASTNAMEn}
% \authormark{LASTNAME et al.}
%
% Use the "et al." form in the case of seven or more authors, or if
% the preferred form is too long to fit in the header.

\authormark{Borkin, Goodman, Halle, \& Alan}

%-----------------------------------------------------------------------
%			Subject Index keywords
%-----------------------------------------------------------------------
% Enter up to 6 keywords describing your paper.  These will NOT be
% printed as part of your paper; however, they will be used to
% generate the subject index for the proceedings.  There is no
% standard list; however, you can consult the indices for past ADASS
% proceedings (http://adass.org/adass/proceedings/).

%\keywords{AXAF, AIPS++, deconvolution, library: digital, Perl,
%          software: architecture}

\keywords{3D Slicer, OsiriX, imaging: three-dimensional, tools}

\begin{abstract}          % Leave intact
The AstroMed project at Harvard University's Initiative in Innovative Computing (IIC) is working on improved visualization and data sharing solutions applicable to the fields of both astronomy and medicine. The current focus is on the application of medical imaging visualization and analysis techniques to three-dimensional astronomical data. The 3D Slicer and OsiriX medical imaging tools have been used to make isosurface and volumetric models in RA-DEC-velocity space of the Perseus star forming region from the COMPLETE Survey of Star Forming RegionÕs spectral line maps. 3D Slicer, a brain imaging and visualization computer application developed at Brigham and WomenÕs HospitalÕs Surgical Planning Lab, is capable of displaying volumes (data cubes as a FITS file or a series of image files), displaying slices in any direction through the volume, generating 3D isosurface models from the volume which can be viewed and rotated in 3D space, and making 3D models of label maps (for example CLUMPFIND output). OsiriX is able to generate volumetric models from data cubes and allows the user in real time to change the displayed intensity level, crop the models without losing the data, manipulate the model and viewing angle, and use a variety of projections.

In applying 3D Slicer to $^{12}$CO and $^{13}$CO spectral line data cubes of Perseus, the visualization allowed for a rapid review of over 8 square degrees and 150,000 spectra, and the cataloging of 217 high velocity points. These points were further investigated in three regions (B5, IC 348, and L1448) and all known outflows were detected, 20 points were identified in these regions as possibly being associated with undocumented outflows, 3 points were identified with a shell around B5 IRS4, and 6 points were identified with a possible large shell spanning across the L1448 and NGC 1333 regions of Perseus. Future work for AstroMed includes making 3D Slicer more compatible with astronomical data including more quantitative tools, and improved segmentation algorithms and displays. All IIC developed tools, as well as 3D Slicer and OsiriX, are freely available.
\end{abstract}
\section{Introduction}
\subsection{The AstroMed Project}
Harvard University's Initiative in Innovative Computing (IIC) was created in 2005 and is an interdisciplinary research and development center. The IIC's goal is to foster interdisciplinary science with the use and advancement of computational technology in order to advance data-intensive science.  The Astronomical Medicine Project (``AstroMed'') is a pilot project of the IIC, and is a collaboration between the Harvard-Smithsonian Center for Astrophysics, the Brigham and Women's Hospital Surgical Planning Laboratory, the Martinos Center for Biomedical Imaging, and the Harvard Medical School.  The project's objective is to address common research features in both the fields of astronomy and medicine including imaging, image analysis, and managing globally distributed large data sets.
\subsection{Why visualize in 3D?}
The third dimension in data cubes is often hard to visualize, and is usually left out of the analysis or display.  In star formation, for example, the third dimension is commonly the velocity component which is essential to determining the cloud's structure, core locations and properties, and velocity features including outflows and shells.  Some common visualization techniques include integrated maps, contour maps and overlays, and channel map movies.  There are many tools that allow one to display 3D data in these ways including DS9, Karma, GAIA, and Aipsview.  Unlike these and other astronomical tools which were originally designed for 2D data, 3D Slicer was designed for the display and analysis of 3D data.  Having a way to display the data in three dimensions in order to take into account RA, DEC, and velocity will generate a wealth of information compared to a two dimensional display of the data.
\section{3D Slicer}
\begin{figure}
\epsscale{0.7}
\plotone{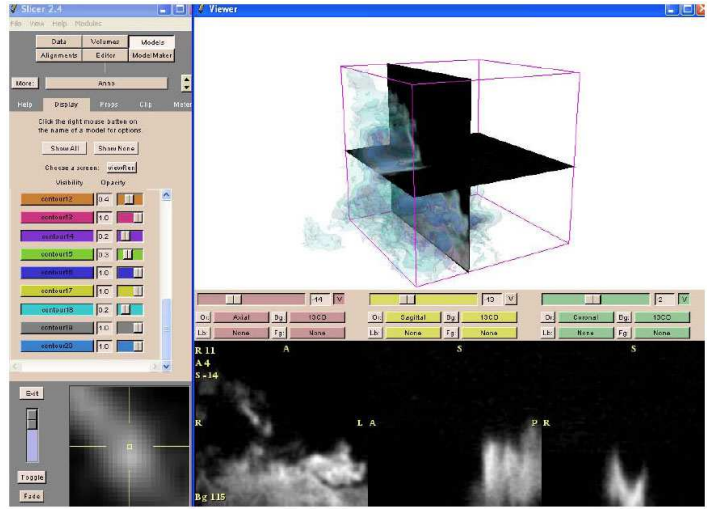}
\caption{The star forming region IC 348 in $^{13}$CO as displayed by 3D Slicer.}
\end{figure}
3D Slicer is a brain imaging and visualization computer application originally developed at the MIT Artificial Intelligence Laboratory and at the Brigham and WomenÕs Hospital's Surgical Planning Lab. It was designed to help surgeons in image-guided surgery, to assist in pre-surgical preparation, to be used as a diagnostic tool, and to help in the field of brain research and visualization (Gering 1999).  3D Slicer is built on ITK and VTK which are C$^{++}$ based programming toolkits developed for the registration, segmentation, and visualization of digital medical imaging data. 3D Slicer was first applied to astronomical data in Borkin et al. (2005) to study the hierarchical structure of star forming cores and velocity structure of IC 348 in $^{13}$CO and C$^{18}$O, and was applied in Borkin (2006) to catalog velocity features in Perseus in order to find previously undetected outflows.

3D Slicer is capable of reading in and displaying volumes (i.e. data cubes in the form of a FITS file or a series of image files), and displaying slices in any direction through a volume.  3D Slicer can easily generate 3D isosurface models from the volume (as shown in Figure 1) which can be viewed and rotated in 3D space (either on a standard monitor or in stereographic with the appropriate monitor or projector).  3D Slicer is also capable of generating 3D models from label maps (for example CLUMPFIND or other algorithmic output).  3D Slicer has built-in segmentation routines and modules which range from a basic watershed to the more complex Expectation-Maximization algorithm.  Images of generated models or slices can be output in the form of individual or series of images.
\section{OsiriX}
\begin{figure}
\epsscale{0.7}
\plotone{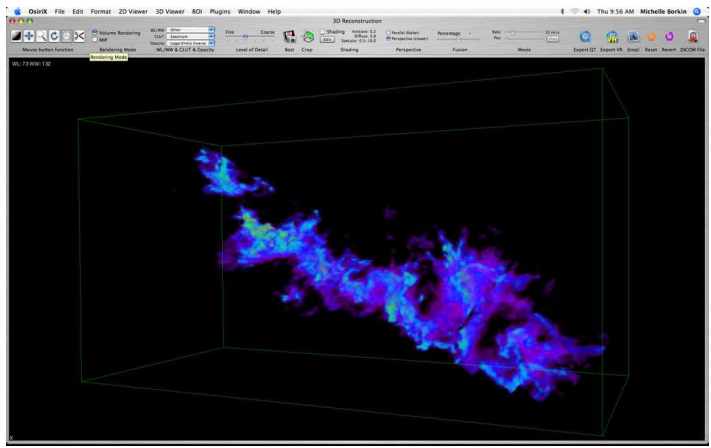}
\caption{The Perseus star forming cloud in $^{12}$CO as displayed in OsiriX.}
\end{figure}
OsiriX is a medical imaging program developed by Dr. Osman Ratib (UCLA) and Dr. Antoine Rosset (University of Geneva) in order to display and analyze large radiological data sets on a desktop computer.  OsiriX, like 3D Slicer, is built on the ITK and VTK programming toolkits and takes advantage of Mac OS X specific graphical and computing capabilities.  OsiriX is able to generate isosurface models and volume rendered models (as shown in Figure 2) from data cubes.  OsiriX allows the user in real time to change the displayed intensity level, crop the models without losing the data, manipulate the model and viewing angle, and use a variety of projections (e.g. maximum intensity projection).  The displayed data can be output as an image, a QuickTime movie, or a QuickTime VR object.  OsiriX's real-time modeling and rendering capabilities make it a strong tool for data exploration.
\section{Astronomical Results}
In applying 3D Slicer to $^{12}$CO and $^{13}$CO spectral line data cubes of Perseus, the visualization allowed for a rapid review of over 8 square degrees and 150,000 spectra, and the cataloging of 217 high velocity points. These points, which were identified by eye in 3D Slicer, were further investigated in three regions (B5, IC 348, and L1448) and all known outflows were detected, 20 points were identified in these regions as possibly being associated with undocumented outflows, 3 points were identified with a shell around B5 IRS4, and 6 points were identified with a possible large shell spanning across the L1448 and NGC 1333 regions of Perseus (Borkin 2006).  Conducting a rapid survey this large using standard techniques for outflow identification (e.g. examining individual spectra or position velocity diagrams) would be impractical.
\section{Conclusion}
Research conducted through AstroMed has served as a proof of concept that medical imaging programs and analysis techniques are advantageous to astronomical imaging and research.  Future work for AstroMed includes making 3D Slicer more compatible with astronomical data including more quantitative tools, improved display capabilities, more file output options, the incorporation of more astronomically relevant segmentation and analysis algorithms, and incorporating interoperability with online data bases.  All IIC developed tools, as well as 3D Slicer and OsiriX, are freely available and can be downloaded online.  For more information, go to http://www.iic.harvard.edu.

% Do not place any material after the references section

\end{document}